\documentstyle[aps,pra,floats,epsf]{revtex}
\def\la{\mathrel{\mathpalette\fun <}}
\def\ga{\mathrel{\mathpalette\fun >}}
\def\fun#1#2{\lower3.6pt\vbox{\baselineskip0pt\lineskip.9pt
  \ialign{$\mathsurround=0pt#1\hfil##\hfil$\crcr#2\crcr\sim\crcr}}}

\begin{document}
\draft

\title{Can a Large Neutron Excess Help Solve the Baryon Loading Problem\\
in Gamma-Ray Burst Fireballs?}
\author{George M. Fuller, Jason Pruet and Kevork Abazajian}
\address{Department of Physics, University of California,
San Diego, La Jolla, California 92093-0319}
\date{April 21, 2000}

%For two column output uncomment following line:
\twocolumn[\hsize\textwidth\columnwidth\hsize\csname@twocolumnfalse\endcsname

\maketitle

\begin{abstract}
We point out that the baryon-loading problem in Gamma-Ray Burst (GRB)
models can be amelioriated if a significant fraction of the baryons
which inertially confine the fireball are converted to neutrons.  A
high neutron fraction in some circumstances can result in a reduced
transfer of energy from relativistic light particles in the fireball
to baryons.  The energy needed to produce the required relativistic
flow in the GRB is consequently reduced, in some cases by orders of
magnitude. This could be relevant to GRB models because a high
neutron-to-proton ratio has been calculated in neutron star-merger
fireball environments. Significant neutron excess also could occur
near compact objects with high neutrino fluxes.
\end{abstract}
\draft

%For two column output uncomment following line:
]

In this Letter we show how the baryon loading problem can be
alleviated in certain gamma-ray burst (GRB) models when significant
numbers of baryons are converted to neutrons. Interestingly, many
of the proposed GRB ``central engines'' involve compact objects which
are themselves highly neutronized, or which are accompanied by intense
neutrino fluxes.  Weak interactions induced by these neutrino fluxes
can result in significant proton-to-neutron conversion, especially if
resonant neutrino flavor transformation takes place
\cite{qfmww93,cfq,mfbf}.

Inferences of the energetics and spectral observations of GRBs imply
(i) total energies in gamma-rays approaching $10^{53}$ ergs for the
most energetic events (in the absence of beaming), and (ii) large
Lorentz factors of the progenitor fireball ($\gamma\sim 10^3$) (for a
recent review, see Ref. \cite{piranrev}). Excessive baryon pollution of
the fireball precludes attainment of these features for many GRB
models. This is a consequence of the conversion of radiation energy in
the electron/positron/photon fireball to kinetic energy in baryons
\cite{sp,kps}.  However, the relatively small cross sections
characterizing the interactions of neutrons with the
electron/positron/photon plasma may afford a solution to this problem.

This can be seen by considering the fictitious limit of completely
noninteracting neutrons.  Imagine that protons inertially tether an
electron/positron/photon fireball via photon Thomson drag on
$e^{\pm}$, which in turn influences protons through Coulomb
interactions. If these protons were suddenly converted to
non-interacting ``neutrons'', then the fireball would expand
relativistically, leaving behind the baryonic component.  Real
neutrons can approximate this limit as they interact with the
electron/positron/photon plasma only via the neutron magnetic dipole
moment.  These cross sections are small compared to the Thomson cross
section $\sigma_{\rm T}$: neutron-electron (positron) scattering has
$\sigma_{{\rm n} e} \sim 10^{-7} \sigma_{\rm T}$ \cite{ahs87}; and
neutron-photon scattering has $\sigma_{{\rm n}\gamma} \sim 10^{-12}
\sigma_{\rm T}$ \cite{gould}.

However, the real limit on the efficacy of this mechanism is the
strong interaction neutron-proton scattering which will dominate the
energy transfer process when conversion of neutrons to protons is
incomplete. Therefore, the degree to which the baryon loading burden
can be lifted in our proposed mechanism will depend on the neutron
excess in the fireball environment.  Here we will measure the
neutron content of the plasma in terms of the electron fraction $Y_e$,
the net number of electrons ($n_{e^-}-n_{e^+}$) per baryon, or
in terms of the neutron-to-proton ratio $Y_e= 1/({\rm n/p}+1)$.

We note that although previous studies have invoked neutrino
oscillations to attempt a baryon loading problem solution
\cite{kluzniak,volkaswong}, none has exploited the $Y_e$-changing
aspect of the weak interaction.

To go beyond the simplistic picture of non-interacting neutrons, we
can consider a two-component ((i) neutrons, and (ii)
protons/$e^\pm$/photons) plasma in the context of a homogeneous
fireball with initial radius, temperature, Lorentz factor, and
electron fraction, $R_0, T_0, \gamma_0$ \& $Y_{e0}$,
respectively. Numerical and analytic work have shown the following
simple scaling laws for such a configuration \cite{kps,spn}:
\begin{eqnarray}
\label{scale}
&\text{For $R<\eta R_0/\gamma_0$} \Rightarrow &\; \cases{\gamma = \gamma_0
\ R/R_0 \cr T = T_0\ R_0/R\cr},\cr\cr
&\text{For $R>\eta R_0/\gamma_0$} \Rightarrow &\; \gamma = \eta.
\end{eqnarray}
As a matter of convenience we will take the scaling to be such that
$\gamma_0=1$. In Eq.\ (\ref{scale}), the ratio of energy in radiation
$E$ to total baryon rest mass $M$ is $\eta \equiv E/M$.

One can relate $\eta$ to the entropy per baryon, $s$, using the number
density of baryons $N={\rho_{\rm b}/{m_p}} = {\rho_{\rm
rad}/{m_p \eta}}$ where $\rho_b$ the baryon component rest mass energy
density and $m_p$ is the proton rest mass.  Using this 
relation, and noting that in terms of the proper entropy
density $S$, the entropy-per-baryon is $s = S/N$, the relation between
$s$, $\eta$, and the temperature $T_0$ is
$s \approx 1250\; \eta ({1~\rm MeV}/{T_0})$.

For a large enough $\eta$, baryon loading is unimportant \cite{sp}.
In fact when
$\eta \ga 10^5 ({E}/{10^{52}~\rm{erg}})^{1/3}
({10^{7}~\rm{cm}}/{R_0})^{2/3},$ 
the fireball becomes optically thin before transferring its energy to
kinetic energy in baryons (here $E$ is the total energy of the fireball).

Written in terms of the time $t$ as measured in a frame
comoving with the fireball the above relations imply
\begin{equation}
\label{scale2}
R = R_0\, e^{t/\tau_{\rm dyn}},\quad
\gamma =  e^{t/\tau_{\rm dyn}},\quad
T = T_0\, e^{-t/\tau_{\rm dyn}},
\end{equation}
for $\gamma<\eta$. Here the dynamic timescale is defined to be the
initial light crossing time, 
$\tau_{\rm dyn} \equiv {R_0/{c }}$. In fireballs resulting
from neutron star mergers, for example, $\tau_{dyn} \sim 2.7 \times
10^{-5} s$, corresponding to an $R_0$ of 8 km \cite{salmwilson}.

A particle co-moving with the expanding plasma experiences a
4-acceleration $a^\mu$, with magnitude $\sqrt{a^\mu a_\mu} =
d\gamma /dR$. As noted above, the force that drags the neutrons along with the
expanding plasma arises princiapally from n-p collisions.  The relative
contribution to the total force on the neutrons from collisions with
electrons and positrons is roughly
$
{F_{n-e}}/{F_{n-p}} \sim
{m_e n_e \sigma_{ne}}/{ m_p n_p \sigma_{np}}
\le 10^{-10} \left({s}/{Y_e}\right)
$
and is small for the conditions we consider. The neutron-photon cross
section is small enough ($\sigma_{n-\gamma} \sim 10^{-36}
[E_\gamma/(1\ \rm MeV)]^2\ cm^2$ where $E_\gamma$ is the photon
energy in the neutron rest frame \cite{gould}) that n-$\gamma$
interactions are negligible.

The relations in Eqs.\ (\ref{scale}) imply that an inertial observer
with time coordinate $t'$ initially (at $t'=0$) comoving with the
plasma sees the plasma accelerate according to $\gamma v =
t'/\tau_{\rm dyn}$.  Hereafter we adopt natural units where $c=1$. If
we denote by $\tau_{\rm coll}^{-1}$ the frequency of neutron/proton
collisions (per neutron), we expect that the two components of the
plasma will achieve a relative velocity given by $v_{\rm rel} \approx
2 {\tau_{\rm coll}}/{\tau_{\rm dyn}}$, where the factor of 2 arises
from the approximate angle independence of the neutron-proton
scattering cross section and the near equality of the neutron and
proton masses. An equivalent expression is found if one considers the
force on the neutrons from collisions with protons \cite{derishev}.
It is clear then that when $\tau_{\rm{dyn}} \gg \tau_{\rm{coll}}$ the
neutrons are coupled to the rest of the plasma. However, decoupling
occurs as these two timescales become comparable. Since the baryon
number density in the plasma frame decreases as $e^{-3 t/\tau_{\rm
dyn}}$, decoupling will occur quickly, {\it i.e.} on a timescale
shorter than $\tau_{\rm dyn}$.

When significant decoupling occurs we can neglect the thermal
contribution to the collision frequency and write
\begin{equation}
\tau_{\rm coll}^{-1} \approx (9\times 10^{12}~{\rm s^{-1}}) {T^3_{\rm
MeV}\over s_5}\ v_{\rm rel}\  Y_e \sigma_{10}\, ,
\end{equation}
where $s_5\equiv s/10^5$ and $\sigma_{10}$ is the neutron-proton
cross-section in units of 10 fm$^2$, and $T_{\rm MeV}$ is the
temperature in MeV. As the precise energy dependence of $\sigma_{10}$
is not important here, it suffices to note a few representative
values: $\sigma_{10}(v_{\rm rel}=0.1) \approx 17$, $\sigma_{10}(v_{\rm
rel}=0.3) \approx 2$ and $\sigma_{10}(v_{\rm rel}=0.6) \approx 0.4$
\cite{pdg}.

The requirement of a non-negligible relative velocity then gives the
decoupling time $t=t_{\rm dec}$ as
\begin{equation}
\label{tdecouple}
{t_{\rm dec} \over \tau_{\rm dyn}} = 4.6 + (1/3)\ln\left({\sigma_{10}
\tau_{-6} T_{0,\rm{MeV}}^3 Y_e\over s_5}\right).
\end{equation}
In the above, $\tau_{-6}\equiv \tau_{\rm dyn}/10^{-6}~\rm sec$, and
$\tau_{\rm coll}$ was evaluated at a terminal velocity of $0.5$. The
calculated decoupling time is logarithmically sensitive to this
choice. In reality the neutrons do not sharply decouple but continue
to interact with the plasma over roughly a dynamical timescale. In
this sense, the $t_{\rm dec}$ appearing in Eq.~(\ref{tdecouple}) is an
``effective'' decoupling time. An accurate determination of $t_{\rm
dec}$ requires solving in detail the neutron and proton transport
equations. However, because of the exponential decrease of density
with time in the plasma frame, the number 4.6 appearing in
Eq.~(\ref{tdecouple}) is only uncertain to approximately $\pm (1/3)$.

Once the neutrons decouple they will have an energy $\gamma_{\rm
dec} (1-Y_e) M$.  The ratio of kinetic energy in neutrons to the
total energy 
in the fireball is then
\begin{equation}
\label{fn}
f_n \approx (1-Y_e) { e^{t_{\rm dec}/\tau_{\rm dyn}}\over \eta} 
\approx 1.3\; (1-Y_e) \left({Y_e \sigma_{10} \tau_{-6}
\over s_5^{4}}\right)^{1/3}.
\end{equation}
 (Here ``total'' energy includes both the thermal $e^{\pm}
/\gamma$ energy and the bulk kinetic energy of baryons.)
From this we see that for $Y_e$ less than 
\begin{equation}
Y_{e,\rm crit} \equiv 0.46 \left({s_5^4\over
\sigma_{10} \tau_{-6}}\right)
\end{equation}
at the time of decoupling the baryon loading problem is diminished.
(Protons and neutrons each move with $\gamma_{\rm dec}$ at the
decoupling point; thereafter, protons will possess larger Lorentz
factors than do average neutrons).  If the plasma remains optically
thick to radiation until the energy in radiation is converted to kinetic
energy of the remaining protons, energy conservation gives the final
Lorentz factor of the protons
\begin{equation}
\label{finalgamma}
\gamma \approx \eta \left({1 - f_n \over Y_e}\right).
\end{equation}
A simple ansatz for the condition that the fireball remains optically
thick after decoupling is $\eta \la (Y_e) 10^5 ({E}/{10^{52}~\rm{erg}})^{1/3}
({10^{7}~\rm{cm}}/{R_0})^{2/3}$. This is obtained by applying the result from
\cite{sp} and making the replacements $\eta \rightarrow (\eta/Y_e)(1-f_n),\,
E \rightarrow (1-f_n) E$ and $s \rightarrow s/Y_e$. 
Note that even for modest values of $f_n$, $\gamma$ can be increased
significantly if $Y_e$ is low.

\begin{figure}
\centerline{\epsfxsize 3in\epsfbox{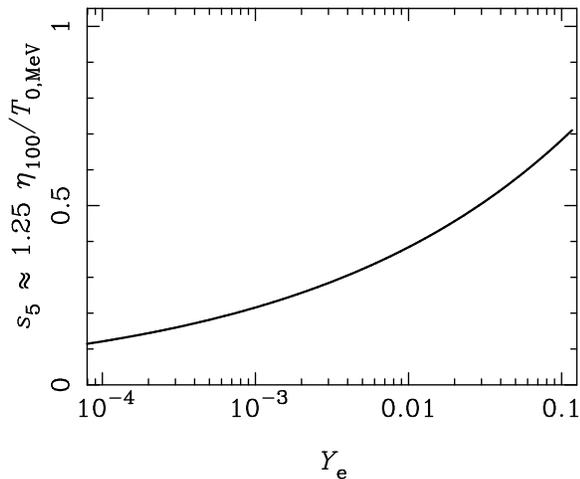}}
\caption[]{
\label{fig}
\small The smallest entropy per baryon, $s_5\equiv s/10^5$, for which
decoupling occurs as a function of $Y_e$. We have taken $\tau _{-6}
\sigma_{10} = 1$.}
\end{figure}

The above results are summarized in Fig. \ref{fig}, where we have
plotted the smallest entropy $s_5$ ($\approx 1.25~\eta_{100}/{T_0}$)
for which decoupling occurs as a function of $Y_e$. 
For example, if
$s_5 = 0.6, T_0 = 2~{\rm MeV}$ (corresponding to $\eta\approx 100$),
and $Y_{e0} = 0.02$, then the final Lorentz factor of the plasma after
neutron decoupling (Eqs.\ (\ref{fn}), (\ref{finalgamma})) would be
$\gamma \approx 1500$, which is 15 times larger than the standard case
of $\gamma=\eta$. As another example, consider the
Ref.\cite{salmwilson} values of $\tau_{-6}=27$ and $Y_{e}=0.1$ and
suppose that $T_0 = 10\,{\rm MeV}$ and $s_5=2.5$ (corresponding to
$\eta=2000$).  In this case we find $\gamma=1.1\times 10^4$, an
increase by a factor of 5.7. Clearly, the importance of this effect
depends on how low $Y_e$ can be.

Two conditions must be met in order to achieve a low $Y_e$ at the time
of decoupling: (i) $Y_e$ must be low initially and (ii) $Y_e$ must not
be unacceptably raised during the evolution of the fireball. We can
divide up the discussion of $Y_e$ in this way because the initial
electron fraction depends in detail on the GRB central engine, whereas
the later evolution of the fireball is generically given by the
relations in Eq.\ (\ref{scale2}). 
%We confine the discussion of the 
%initial electron fraction ($Y_{e0}$) to environments where neutrino
%heating is important, both because of the existence of extensive
%studies of the relation between electron fraction and neutrino
%parameters in such systems and because it is easiest to see how a low
%$Y_{e0}$ might be obtained in these environments.  In fact, 
Many proposed GRB central engines involve neutrino heating or are
sited in environments subject to intense neutrino fluxes
\cite{salmwilson,Paczynski90,Goodman86,Paczynski86,MesRees92,woosley93,macwoos,brown,smogrb}.
General discussions of the relation between neutrino processes and the
dynamics of outflow may be found in
Refs. \cite{salmwilson,dsw,qw,cardall}. However, the details of neutron
decoupling are insensitive to how $Y_e$ is set and we are not arguing
for a specific GRB site.

The processes which have a significant effect on $Y_e$ in the fireball
environment are lepton capture/decay involving free nucleons and
inelastic ${\rm nn}\rightarrow {\rm np}\pi$ scattering (charged
pion-nucleon bremsstrahlung),
\begin{mathletters}
\begin{eqnarray}
\label{nun} \nu_e + \rm{n} &\rightleftharpoons&  \rm{p} + e^- \\ 
\label{nup} \bar\nu_e + \rm{p} &\rightleftharpoons&  \rm{n} + e^+ \\
\label{ndecay} {\rm n} &\rightarrow& {\rm p} + e^-+\bar\nu_e \\
\label{pis} {\rm n}+{\rm n} &\rightarrow& {\rm n}+{\rm p}+\pi^-.
\end{eqnarray}
\end{mathletters}
In general, $Y_e$ is set by the competition between the above
processes \cite{qfmww93,qf95}.  For the range of fireball parameters
of interest to us, free neutron decay (\ref{ndecay}) is unimportant as
the fraction of neutrons decaying during the evolution of the fireball
is $\sim 10^{-9}\tau_{-6} \, \ln{\eta} $.  Furthermore, as lepton capture
is only important during the early, hot, evolution of the fireball and
inelastic nucleon-nucleon scattering only occurs after neutron
decoupling, the lepton capture and pion bremsstrahlung processes may
be considered separately.

In environments where neutrino heating is important the forward
reactions (\ref{nun}) and (\ref{nup}) can dominate in setting the
electron fraction \cite{qfmww93}.  Integration of the rate equations
corresponding to the lepton capture processes gives ${\rm n/p} \approx
\lambda_{\bar\nu_e\rm p}/\lambda_{\nu_e\rm n} \approx
(L_{\bar\nu_e}\langle E_{\bar\nu_e}\rangle)/ (L_{\nu_e}\langle
E_{\nu_e}\rangle)$, where $\lambda_{\bar\nu_e\rm p}$ and
$\lambda_{\nu_e\rm n}$ are the rates for the reactions in Eqs.\
(\ref{nun}) and (\ref{nup}), $\langle E_{\bar\nu_e}\rangle$ and
$\langle E_{\nu_e}\rangle$ are the average energies characterizing the
energy spectra of the $\bar\nu_e$ and $\nu_e$ neutrinos, respectively,
while $L_{\bar\nu_e}$ and $L_{\nu_e}$ are the corresponding energy
luminosities.  Absent neutrino oscillations and flavor/type mixings,
any thermal neutrino emission scenario from a compact object will
yield a characteristic average neutrino energy heirarchy for solar
mass scale objects: $\langle E_{\nu_\mu}\rangle \approx \langle
E_{\bar\nu_\mu}\rangle \approx \langle E_{\nu_\tau}\rangle \approx
\langle E_{\bar\nu_\tau}\rangle > \langle E_{\nu_{\bar e}}\rangle >
\langle E_{\nu _e}\rangle$. These considerations are consistent with
findings in Ref.\ \cite{salmwilson} in which a hard $\bar\nu_e$
spectrum from a collapsing neutron star leads to an electron fraction
in the fireball of $Y_e \sim 0.1$.

If the $\nu_e$ component of the neutrino emission were to disappear or
be greatly reduced, then the competition inherent in the above
equations would be unbalanced in favor of the reaction $\bar\nu_e +
{\rm p \rightarrow n} + e^+$.  This, in turn, would result in the
wholesale production of neutrons.  In fact, several schemes involving
matter-enhanced active-sterile neutrino transformation have been
proposed as a way of enabling $r$-process nucleosynthesis in
neutrino-heated supernova ejecta: one of these involves
matter-enhanced $\nu_e \rightleftharpoons \nu_s$ and $\bar\nu_e
\rightleftharpoons \bar\nu_s$ \cite{cfq}; the other involves
matter-enhanced conversion $\nu_{\mu,\tau} \rightleftharpoons \nu_s$
followed by an active-active matter-enhanced conversion
$\nu_{\mu,\tau} \rightleftharpoons \nu_e$ \cite{mfbf}. In either case,
the intial $\nu_e$ flux can be reduced by more than an order of
magnitude and, in turn, this can translate into a substantial decrease
in the initial $Y_e$. (Just how low depends on central engine outflow
hydrodynamics and on neutrino background effects
\cite{mfbf,qf95,patelfuller,mclaughlin}.)

If we demand that an initially low $Y_e$ not be raised above
$Y_{e,{\rm crit}}$, consideration of lepton capture on neutrons allows
us to place rough constraints on the fireball and neutrino parameters.
We incorporate the uncertainty in the initial fireball evolution by
supposing that the relations in Eq.\ (\ref{scale2}) are valid only
after the fireball has a Lorentz factor $\gamma_{i}$ and temperature
$T_i$. Consideration of positron capture after $\gamma=\gamma_i,\,
T=T_i$ then leads to $T_i < (22~{\rm MeV}) \left(Y_{e,\rm
crit}/\tau_{-6}\right)^{1/5}$ Similarily, consideration of $\nu_e$
capture on neutrons leads to $ T_{\nu_e} < (40~{\rm MeV}) \gamma_i
\left({Y_{e,\rm crit}}/{\tau_{-6}}\right)^{1/5}.$ In deriving this
limit we have taken the $\nu_e$ spectrum to be a Fermi-Dirac blackbody
with temperature $T_{\nu_e}$ and zero chemical potential. This limit
could be modified or weakened if $\nu_e$ flavor transformation occurs.

\begin{figure}
\centerline{\epsfxsize 3in\epsfbox{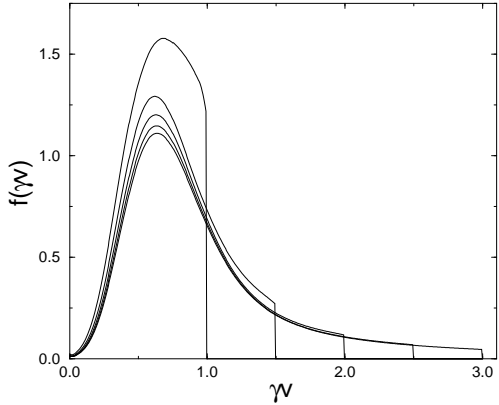}}
\caption[]{
\label{fig2}
\small The neutron distribution function $f(\gamma v)$ (normalized so that
$\int{f d(\gamma v) = 1}$) at several time slices as calculated under
the assumptions given in the text. The different time slices correspond
to a plasma $\gamma v$ of 1, 1.5, 2, 2.5, 3, as measured in a frame
comoving with the plasma near the decoupling point.}
\end{figure}

Determining the increase in $Y_e$ due to pion production requires a
proper treatment of neutron transport in the plasma. However, an upper
limit on the increase is readily obtained by considering the extreme
case where (i) the protons are frozen into the accelerating plasma (ii)
non-forward n-p collissions are assumed to result in maximal momentum
exchange (iii) n-n collissions are ignored except as a post-processing
step to determine $\pi$ production and (iv) the change in $Y_e$ due to
inelastic n-p and inelastic p-p scatterings is ignored. This simple
picture gives an upper limit on the increase in $Y_e$ because an
exchange of any of the assumptions (i)-(iii) for more realistic ones
has the effect of decreasing the velocity dipersion of the neutrons.
By a calculation with the above assumptions we obtain the upper limit
on the increase in $Y_e$ to be $\Delta Y_e \la 10^{-3}/Y_{e0}$, where
$Y_{e0}$ is the initial electron fraction. Fig. (\ref{fig2}) displays
the evolution of the neutron distribution function as calculated with
the above assumptions. The distribution function drops sharply at the
instantaneous plasma velocity because there is no mechanism for
boosting neutrons to higher velocity.

The ineffectiveness of pion bremsstrahlung in increasing $Y_e$ may be
attributed to the fact that the processes which increase $Y_e$ result
from a two step process (i.e. an n-p scattering boosts a neutron which
then inelastically scatters with another neutron), the decrease of
$\sigma(v) \, v$ with increasing velocity (the product of cross
section and relative velocity decreases by a factor of 5 as $v$
increases from 0 to c \cite{pdg}), and the late onset of this process
in n-n scattering. Pion production does not begin until the pion mass
threshold is reached at a relative velocity of $0.645 $ and even at a
relative velocity of $0.728 $ (center of mass energy 2.08 GeV), the
inelastic contribution to the cross section is only 8\% of the total
\cite{pdg}. For $\Delta Y_e > Y_{e0}$ our perturbative approach to
calculating $\Delta Y_e$ breaks down and the increase in $Y_e$ may
cause a recoupling of the proton and neutron flows. Note that our
calculation only gives an upper limit on the increase in $Y_e$ due to
inelastic n-n scattering. A more careful transport calculation will
likely show a smaller increase in $Y_e$.  It is consistent then to
discuss decoupling at low $Y_e$ for a wide range of fireball
parameters.  The final Lorentz factor of the plasma may then be
substantially increased for a given energy input and baryon load.

Aside from the fireball energetics versus $Y_e$ issue addressed here,
neutron-proton separation recently has been shown to have implications
for the electromagnetic and high-energy neutrino signatures of GRBs
\cite{derishev,waxbah97,bahcall00}.  Future large volume detectors
such as AMANDA/ICECUBE will be able to provide high energy neutrino
data on GRBs \cite{halzen}.  Perhaps the details of neutron decoupling
and the associated electromagnetic/neutrino signature could allow a
diagnostic of the weak interaction physics deep in GRB central engine
environments.

It is a pleasure to acknowledge discussions with N.~Dalal, M.~Patel, X.~Shi
and J.~R.~Wilson.  This research was supported in part by NSF Grant
PHY98-00980, an IGPP grant, and a NASA GSRP for KA.

\end{document}